# Phase modulated pulse interferometry for simultaneous multi-channel ultrasound detection

**YOAV HAZAN**[1,*] **AND AMIR ROSENTHAL**[1]

[1]*Technion - Israel Institute of Technology, Haifa, 3200003, Israel*
*\*Corresponding author: yoav.hazan@campus.technion.ac.il*

**In optical detection of ultrasound, resonators with high Q-factors are often used to maximize sensitivity. However, in order to perform parallel interrogation, conventional interferometric techniques require an overlap between the resonator spectra, which is difficult to achieve with high Q-factor resonators. In this paper, a new method is developed for parallel interrogation of optical resonators with non-overlapping spectra. The method is based on a phase modulation scheme for pulse interferometry (PM-PI) and requires only a single photodetector and sampling channel per ultrasound detector. Using PM-PI, parallel ultrasound detection is demonstrated with four high Q-factor resonators. © The Authors**

Piezoelectric (PE) transducers represent a fundamental technology in the field of ultrasound (US), which has been essential to the development of medical sonography [1]. The success of PE technology may be attributed to its capability to generate and detect US with a single device and the availability of US arrays whose elements may be accessed simultaneously. With the recent emergence of hybrid imaging modalities in which the generation of US is performed via the absorption of electromagnetic (EM) energy in the tissue, many imaging systems have been developed that use PE transducers only in detection mode [2–7]. In the field of optoacoustic tomography, in which light is used to generate US, the availability of US arrays has led to the development of imagers capable of producing 2D and 3D images in real time [4,5].

Notwithstanding recent progress, many challenges remain in the development of hybrid imaging modalities that stem from the limitations of PE technology. PE transducers are opaque, susceptible to EM interference, and often have limited bandwidths and acceptance angles that lead to image artifacts [6]. In addition, the miniaturization of PE arrays generally leads to loss of sensitivity, limiting minimally invasive applications [7].

One of the alternatives to PE technology for US detection in hybrid-imaging applications is optical interferometry. Interferometric detectors of US offer immunity to EM interference [3], may be produced from transparent materials [8–12], and can achieve wide bandwidths and acceptance angles [10,13,14]. In addition, when optical resonators with high Q-factors are employed, the detector may be miniaturized without compromising sensitivity [13,15]. When acoustic waves impinge on an optical resonator, they modulate its refractive index owing to the elasto-optic effect and deform its structure, leading to a modulation in the resonance wavelength [16].

Conventionally, optical resonators are interrogated with CW lasers tuned to their resonance; where US induced resonance shifts are translated to intensity variations at the output of the resonator. While arrays of optical resonators may be produced in a single platform [17–19], parallel interrogation with a single CW laser is challenging since it requires a good overlap between the spectra of all the resonators. In addition, external disturbances like vibrations and bending may shift the resonances away from each other, preventing their simultaneous interrogation with a single CW laser; in the case of severe mechanical disturbances, even serially tuning the laser to each resonator separately may be technically challenging [20]. While the challenge of parallelization may be mitigated by using resonators with lower Q-factors [21], or avoided by using schemes that do not rely on resonators [22], these approaches may lead to reduced sensitivity. Recently, Wei *et al.* demonstrated that the robustness of CW methods may be improved if the resonator is integrated in a fiber ring laser [23]; however, parallelization of this approach has yet to be demonstrated.

Pulse interferometry (PI) has been developed as an alternative approach for interrogating optical resonators, which may potentially overcome some of the limitations that have characterized CW interrogation [24–27]. In PI, the source is based on a pulse laser whose bandwidth is sufficiently wide to cover the entire bandwidth in which the resonances may occur. In [25,26], this property was used to achieve a high dynamic range and robust operation under volatile environmental conditions. While PI may in principle be used to interrogate several resonators with non-overlapping spectra with a single source, the demodulation schemes used in [24,25] were not scalable, and would lead to unacceptable cost per channel.

In this paper, we report on a new version of PI that may be scaled to simultaneously interrogate multiple optical resonators with non-overlapping spectra. Our method is based on a phase-modulation scheme performed at the output of the source, which couples between the intensity and wavelength at the output of the resonators. Using a simple demodulation algorithm, US -induced shifts in the resonators' wavelengths may be decoded from intensity measurements. Accordingly, only a single photodetector and sampling channel are required per resonator. Using the proposed phase-modulation PI (PM-PI), we demonstrate, for the first time to our knowledge, parallel interrogation of four resonators with non-overlapping spectra.

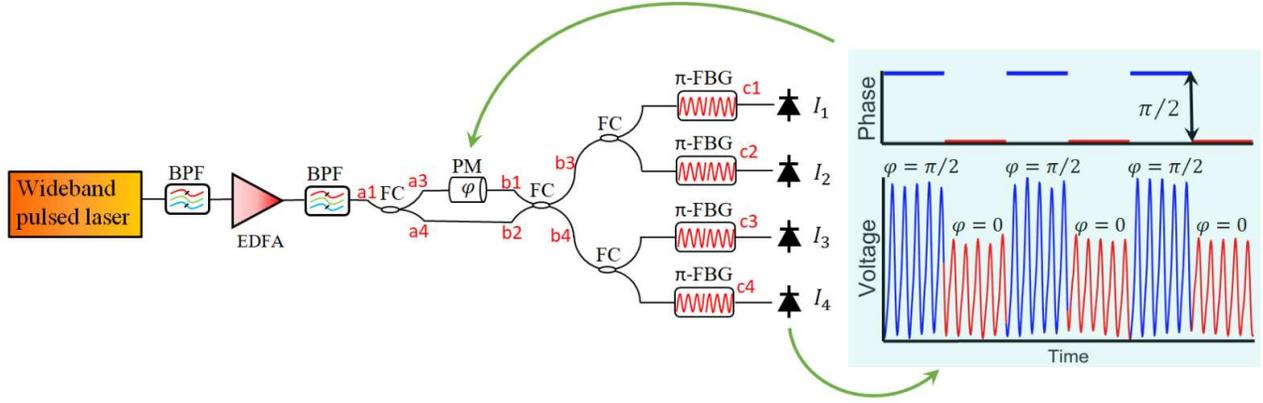

Fig. 1. A schematic drawing of the PM-PI system used in this work to interrogate 4 resonators, implemented with π-phase shifted fiber Bragg gratings (π-FBGs). A wideband pulse laser with band-pass filters (BPFs) and an erbium-doped fiber amplifier (EDFA) create a source with a high spectral power density and sufficient bandwidth to cover the spectra of all the resonators. The modulation unit is an unbalanced Mach-Zehnder interferometer (MZI) with a phase modulator (PM) that is switched between two phase values with a difference of $\pi/2$. For each phase value, the pulses interfere differently at the output of each resonator depending on the phase difference in the MZI for the specific resonance wavelength of that resonator. In the current implementation, the duration of each phase value of the modulator corresponded to 5 pulses of the laser.

An illustration of the PM-PI system used in this work is given in Fig. 1. A wideband pulse laser is employed together with optical band-pass filters (BPFs) and an erbium-doped fiber amplifier (EDFA), which are used to increase the spectral power density of the source while maintaining an acceptable bandwidth. An unbalanced Mach-Zehnder interferometer (MZI) with a phase modulator (PM) on one of its arms is connected at the output of the source; where the phase is switched between two values with a difference of $\pi/2$. The resonators, implemented by π-phase shifted fiber Bragg gratings (π-FBGs) [15], are connected to photodetectors whose voltage signals are sampled. As the analysis in the following shows, the output of each resonator is switched between two interferometric states, which together enable us to monitor wavelength shifts of its resonance.

In our analysis, we assume that the PM has only two phase states that may be experienced by each pulse; whereas the transition between the two states occurs at times when no light passes through the PM. We follow the electric field at different positions in the PM-PI setup, marked with the red letters a-c in Fig. 1. The electric field at the input of the MZI, i.e. point (a1), is given by $e_1^a = 2e(\omega)$; where $\omega$ is the angular frequency. At the two outputs of the first coupler, points (a3) and (a4), the electric fields respectively given by $e_3^a = \sqrt{2}e(\omega)$ and $e_4^a = \sqrt{2}ie(\omega)$; whereas at the two inputs of the second coupler, points (b1) and (b2), one obtains:

$$e_1^b = \sqrt{2}e(\omega)\exp(i\varphi_{PM} + i\omega n l_1 / c) \quad (1a)$$

$$e_2^b = \sqrt{2}ie(\omega)\exp(i\omega n l_2 / c) \quad (1b)$$

where $l_1$ and $l_2$ are the lengths of the top and bottom arms of the MZI, respectively, $n$ is the refractive index of the fiber, $c$ is the speed of light, and $\varphi_{PM}$ is the phase contribution of the PM, which may be equal to either 0 or $\pi/2$. Neglecting global phase accumulation, the electric fields at the output of the second coupler are given by:

$$e_3^b = e(\omega)\left[\exp(i\varphi_{PM}) - \exp(i\omega n \Delta l / c)\right] \quad (2a)$$

$$e_4^b = e(\omega)\left[\exp(i\varphi_{PM}) + \exp(i\omega n \Delta l / c)\right] \quad (2b)$$

where $\Delta l = l_2 - l_1$.

At the final stage of the setup, the electric fields $e_3^b$ and $e_4^b$ are filtered by the resonators, whose field transmission functions we denote by $H_i(\omega)$; where $i = 1, ..., 4$. For each resonator, we denote the central frequency by $\omega_i$ and its bandwidth by $\Delta\omega_i$. We assume that $\Delta l$ is sufficiently small so that $\exp(i\omega n \Delta l / c)$ is approximately constant over $\Delta\omega_i$ ($i = 1, ..., 4$); the corresponding mathematical condition is given by

$$\Delta\omega_i n \Delta l / c \ll 2\pi \quad (3)$$

When Eq. (3) is fulfilled, the electric fields at the output of the resonators may be approximated by the following expressions:

$$e_{1,2}^c = e_{1,2} H_{1,2}(\omega)\left[\exp(i\varphi_{PM}) - \exp(i\omega_{1,2} n \Delta l / c)\right] \quad (4a)$$

$$e_{3,4}^c = e_{3,4} H_{3,4}(\omega)\left[\exp(i\varphi_{PM}) + \exp(i\omega_{3,4} n \Delta l / c)\right] \quad (4b)$$

where $e_i = e(\omega_i)$. The power at the output of each π-FBG is thus given by

$$P_i^c = P_i\left[1 - \cos(\varphi_i + \varphi_{PM} - \omega_i n \Delta l / c)\right] \quad (5)$$

where $P_i = 2\int |e_i H_i(\omega)|^2 d\omega$, and where $\varphi_i = 0$ for $i = 1, 2$ and $\varphi_i = \pi$ for $i = 3, 4$. In practice, $\Delta l$ may not fully fulfill the condition in Eq. (3), leading to the following modification in Eq. (5) [24]:

$$P_i^c = P_i\left[1 - \eta_i \cos(\varphi_i + \varphi_{PM} - \omega_i n \Delta l / c)\right] \quad (6)$$

where $\eta_i \leq 1$ is the visibility of the interference.

Assuming that the two phase states of the PM lead to $\varphi_{PM} = 0$ and $\varphi_{PM} = \pi/2$, two expressions are obtained for the measured power at each channel, corresponding to the two states:

$$P_{i,0}^c = P_i\left[1 - \eta_i \cos\left(\varphi_i - \omega_i n\Delta l / c\right)\right] \quad (7a)$$

$$P_{i,\pi/2}^c = P_i\left[1 + \eta_i \sin\left(\varphi_i - \omega_i n\Delta l / c\right)\right] \quad (7b)$$

When $P_i$ and $\eta_i$ are known from a pre-measurement calibration procedure, the sine and cosine in Eqs. (7a) and (7b) may be readily calculated from $P_{i,0}^c$ and $P_{i,\pi/2}^c$ and used to calculate the phase $\phi_i = \varphi_i - \omega_i n\Delta l / c$ using the four-quadrant inverse tangent (atan2) and a phase unwrapping algorithm [28].

When an acoustic pulse impinges on the resonator, it leads to a modulation in $\omega_i$. Accordingly, we define $\omega_i(t) = \omega_i^{dc} + \omega_i^{ac}(t)$; where $\omega_i^{dc}$ represents the resonance frequency before the arrival of the acoustic pulse, and $\omega_i^{ac}(t)$ is the US-induced perturbation, which we wish to recover. Assuming that the MZI is not exposed to the acoustic pulse, the term $n\Delta l$ may be regarded as constant during the acoustic measurement. Thus, $\omega_i^{ac}(t)$ may be readily recovered from $\phi_i(t)$:

$$\omega_i^{ac}(t) = c\left[\varphi_i - \phi_i(t)\right] / n\Delta l - \omega_i^{dc} \quad (8)$$

In practice, $\omega_i^{ac}(t)$ is obtained by applying a high-pass filter on the right hand side of Eq. (8) while ignoring the values of $\varphi_i$ and $\omega_i^{dc}$.

Figure 2 shows the resonance spectra of the 4 π-FBGs (TeraXion Inc., Quebec, Canada) used in our measurements as a function of detuning from the central wavelength of 1549 nm. The resonances had a full-width-at-half-maximum (FWHM) of approximately 4 pm; whereas the spectral distance between two resonances had a maximum value of approximately 43 pm, preventing parallel interrogation with CW techniques. The pulse laser (M-Comb model, Menlo Systems GmbH, Martinsried, Germany) had a central wavelength of 1560 nm, spectral bandwidth of 40 nm, pulse repetition rate of 250 MHz, pulse duration of approximately 90 fs, and average power of 75 mW, whereas the BPFs had a spectral width of 0.4 nm around the resonance wavelength, sufficiently wide to cover all the 4 resonances shown in Fig. 2. The output power from each of the π-FBGs was approximately 25 μW when the EDFA was set to 100 mW average power output. All the components in our system were implemented with polarization-maintaining fibers to avoid polarization fading in the setup [29].

The MZI in Fig. 1 had an imbalance of $\Delta l = 7$ cm and included a PM with a modulation bandwidth of 20 GHz (PM-5S5-10-PFA-PFA-UV-UL, EOspace). The PM was fed with a square voltage signal with a frequency of 25 MHz and duty cycle of 50% and the modulated signals at the output of the π-FBGs were detected by photodetectors with a bandwidth of 1.5 GHz (DET01CFC, Thorlabs), connected to a 4-channel oscilloscope with a bandwidth of 1.5 GHz (Keysight, DSOX4154A). As shown in Fig. 1, each cycle of the voltage signal corresponded to 10 pulses, half of which with the response given in Eq. 7a and half with the response of Eq. 7b. Since the voltage signal was not synchronized with the repetition rate of the laser, some the laser pulses occasionally overlapped with the transition between the two states of $\varphi_{PM}$. To avoid transition effects, $P_{i,0}^c$ and $P_{i,\pi/2}^c$ were extracted from the photodetector signals by calculating the median value for every 5 pulse peaks within half a cycle of the PM.

The performance of PM-PI was tested using an acoustic setup similar to the one used in [15,24,26]. Briefly, the π-FBGs were placed in a water bath along with an US transducer with a central frequency of 1 MHz. To maximize the resonance frequency shift in the fibers, the orientation of the transducer was adjusted to an angle of 30° with respect to the optical fibers, leading to excitation of a guided acoustic wave in the fibers, which has been previously shown to generate a stronger response than normal-incidence waves [30,31].

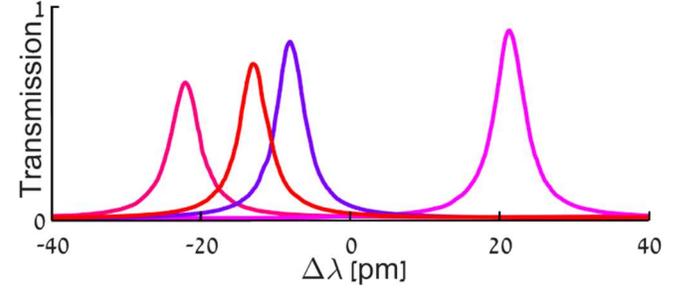

Fig. 2. The transmission spectra of the four different π-FBG resonances used in the system shown in Fig. 1.

In the first measurement, the signal from a single π-FBG was measured using PM-PI and the original implementation of PI developed in [24], in which active stabilization of the MZI was used (A-PI). Since A-PI suffers from non-linear signal folding when the acoustic signal is too strong [26], the magnitude of the US burst was chosen to be sufficiently small to fit the linear-operation range of the MZI, which was 600 MHz in our implementation. Figure 3 shows the frequency shift of the π-FBG, $\Delta\nu(t) = \omega^{ac}(t)/2\pi$, measured with A-PI and PM-PI. The acoustic signals and corresponding spectra measured by the two techniques are in very good agreement as shown in Figs. 3a and 3b, respectively. The small differences may be attributed to a deviation from the π/2 phase difference between the two interleaved signals in the implementation of PM-PI. The measurement bandwidth in PM-PI was limited to 12.5 MHz, i.e. half of the phase-modulation frequency. Using this bandwidth, the minimum detectable resonance shifts for A-PI and PM-PI were 3 MHz and 10 MHz, respectively.

In the second measurement, the ability of PM-PI for parallel interrogation was demonstrated with the four π-FBGs whose spectra are shown in Fig. 2. The π-FBGs were positioned in proximity to each other and in a 30° angle with respect to the US transducer. To demonstrate that, in contrast to A-PI [26], PM-PI does not suffer from signal folding, the acoustic pressure produced by the US transducer was increased by a factor of 4 with respect to the first measurement. The measured resonance shifts of the four π-FBGs are shown in Fig. 4. Since the π-FBGs were placed at different positions within the acoustic beam with possible differences in their exact orientations, the signals vary in their amplitude and delays. The strongest measured resonance shift had a peak-to-peak value of 1 GHZ, i.e. twice larger than the resonance width. Such a large shift cannot be properly measured using A-PI or conventional CW interrogation, as previously demonstrated in [26].

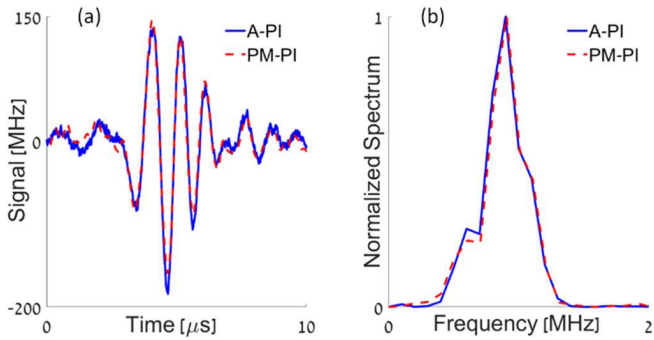

Fig. 3. Optical resonance frequency shifts induced by equivalent US pulses using both active demodulation (A-PI) and phase modulation (PM-PI) techniques. The resonance shifts are presented in time domain (a) and frequency domain (b).

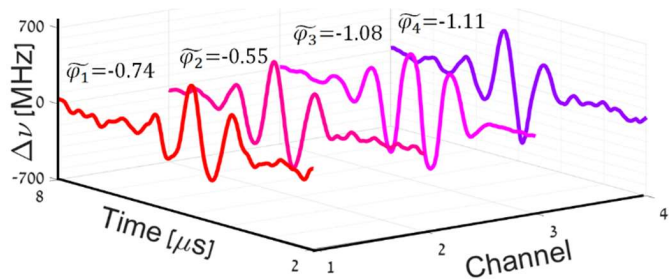

Fig. 4. Optical resonance frequency shifts from 4 different π-FBGs. The four different signals are measured simultaneously by the PM-PI scheme. The mean phase $\tilde{\varphi}_i = \varphi_i - \omega_i^{dc} n \Delta l / c$ is noted for each of the four signals.

In conclusion, we demonstrated a novel scheme for parallel interrogation of resonator-based interferometric detectors of US. Our scheme is based on a variation PI, in which the pulses are modulated at the input of the resonators by a PM. The modulation enables the coding in time of two interference states, which facilitates digital demodulation of the desired signals from simple power measurements. While the PM in this work operated at only 10% of the pulse rate, reducing the measurement bandwidth by the same proportion, in an ideal implementation the PM modulation frequency may be set at 50% of the pulse repetition rate. For the pulse laser used in this work, which had a repetition rate of 250 MHz, PM-PI with a modulation frequency of 125 MHz corresponds to an acoustic bandwidth of 62.5 MHz, which is compatible with most imaging applications.

PM-PI overcomes a major limitation of interferometric detectors of US – the inability to simultaneously interrogate several resonators with non-overlapping spectra. However, the current implementation of PM-PI comes at the price of increased noise in comparison to A-PI owing to the electronics used in the signal modulation. Future implementations of PM-PI will thus require using low-noise signal generators and demodulation electronics to reduce the noise and enable the development of US detector arrays that achieve the miniaturization and sensitivity levels possible with high Q-factor resonators [13,32]. Such US arrays may be used to significantly improve the performance of hybrid imaging systems for which no compatible US-array technology exist [7,33].

**Funding.** Israel Science Foundation (942/15, 694/15); Niedersächsisches Vorab (ZN3172); EVPR fund at the Technion.